\documentclass[12pt]{article}

\usepackage{amssymb}
\usepackage{amsmath}
\usepackage{amsfonts}
\usepackage{graphicx}
\usepackage{mathrsfs}
\usepackage{comment}
\usepackage{cite}

\newcommand{\Slash}[1]{{\ooalign{\hfil#1\hfil\crcr\raise.167ex\hbox{/}}}}

\newcommand{\beq}{\begin{equation}}  \newcommand{\eeq}{\end{equation}}
\newcommand{\bef}{\begin{figure}}  \newcommand{\eef}{\end{figure}}
\newcommand{\bec}{\begin{center}}  \newcommand{\eec}{\end{center}}
\newcommand{\non}{\nonumber}  \newcommand{\eqn}[1]{\beq {#1}\eeq}
\newcommand{\laq}[1]{\label{eq:#1}}  

\newcommand{\Eq}[1]{Eq.(\ref{eq:#1})}

\newcommand{\eq}[1]{(\ref{eq:#1})}
\newcommand{\Sec}[1]{Sec.\ref{chap:#1}}
\newcommand{\ab}[1]{\left|{#1}\right|}
\newcommand{\vev}[1]{ \left\langle {#1} \right\rangle }

\newcommand{\lac}[1]{\label{chap:#1}}
\newcommand{\SU}[1]{{\rm SU{#1} } }

\def\({\left(}
\def\){\right)}

\def\O{\mathcal{O}}
\def\U{\mathop{\rm U}}

\def\ebq{\eeq \beq}
\newcommand{\OR}{~{\rm or}~}
\newcommand{\AND}{~{\rm and}~}
\newcommand{\EV}{ {\rm ~eV} }

\newcommand{\MEV}{ {\rm ~MeV} }
\newcommand{\GEV}{ {\rm ~GeV} }
\newcommand{\TEV}{ {\rm ~TeV} }
\def\o{\over}

\def\d{\delta}

\def\f{\phi}

\def\h{\theta}
\def\k{\kappa}
\def\l{\lambda}
\def\m{\mu}
\def\n{\nu}
\def\p{\psi}

\def\s{\sigma}

\def\D{\Delta}
\def\G{\Gamma}

\def\L{\Lambda}
\def\F{\Phi}

\def\ol{\overline}
\def\tl{\tilde}
\def\*{\dagger}

\begin{document}
\begin{titlepage}
\begin{center}

\vspace{2.5cm}

{\Large\bf Dark Matter, Neutrino mass, Cutoff for Cosmic-Ray Neutrino, and Higgs Boson Invisible Decay from a Neutrino Portal Interaction}
\vspace{1.5cm}

{ {\bf  Wen Yin \footnote{email: wyin@ihep.ac.cn, yinwen@kaist.ac.kr}}}

  {\em IHEP, Chinese Academy of Sciences, Beijing 100049,  China}\\ 
 { \em Department of Physics, KAIST, Daejeon 34141, Korea}

\vspace{1.5cm}
\abstract
We study an effective theory beyond the standard model (SM) where either of two 
additional gauge singlets, a 
Majorana fermion and a real scalar, constitutes all or some fraction of dark matter. In particular, we focus on the masses of the two singlets in the range of $\O(10)\MEV-\O(10)\GEV$, with a neutrino portal interaction which plays important roles not only in particle physics but also in cosmology and astronomy.
We point out that the dark matter abundance can be thermally explained with (co)annihilation, where the dark matter with a mass greater than 2 GeV can be tested in future lepton colliders, CEPC, ILC, FCC-ee and CLIC, in the light of the Higgs boson invisible decay.
 When the gauge singlets are lighter than $\O(100)\MEV$, the interaction can affect the neutrino propagation in the universe due to its annihilation with the cosmic background neutrino into the gauge singlets.  Although can not be the dominant dark matter in this case, the singlets are produced by the invisible decay of the Higgs boson at a rate fully within the reach of the future lepton colliders.
In particular, a high energy cutoff of cosmic-ray neutrino, which may account for the non-detection of Greisen-Zatsepin-Kuzmin (GZK) neutrinos or non-observation of Glashow resonance, can be set. 
Interestingly, given the cutoff and the mass (range) of the WIMP, a neutrino mass can be ``measured" kinematically.

\end{center}
\end{titlepage}
\setcounter{footnote}{0}
\section{Introduction}
Weakly Interacting Massive Particles (WIMPs) are promising candidates of dark matter\cite{Arcadi:2017kky, Bertone:2004pz, Bergstrom:2000pn, Jungman:1995df}. 
However, the WIMPs with mass $6\GEV -\O(10^2)\TEV$ has been severely constrained by the XENON, LUX and PandaX experiments \cite{Akerib:2013tjd, Xiao:2014xyn, Akerib:2015rjg, Akerib:2016vxi, Tan:2016diz, Fu:2016ega, Aprile:2017iyp, Aprile:2016swn}.
This situation gives the motivation to investigate WIMPs lighter than GeVs. 
Such a WIMP should be a singlet of the standard model (SM) gauge group to avoid the LEP constraints \cite{Olive:2016xmw}. If a gauge singlet dark matter is stabilized by a hidden symmetry, its possible interaction with the SM particles is represented by a portal coupling $\O_{SM} \O_{DM}$, where $\O_{SM}$ ($
 \O_{DM}$) is a SM gauge singlet operator composed only of the SM fields (only of the hidden fields including the WIMP).

It is interesting to study a neutrino portal interaction, i.e.
$\O_{SM}=\f_{H}\cdot L$, where $L$ is a Weyl spinor for a left-handed lepton and we will 
take Weyl representation hereafter; $\f_H$ is the Higgs doublet field; the dot denotes the 
contraction of the $\SU(2)$ gauge indices, while the Lorentz indices are omitted. 
This is because this interaction can be not only a window of the SM to a dark sector but also affects neutrino and the Higgs boson physics. 

Neutrino portal dark matter has been studied in several contexts: 
asymmetric dark matter \cite{Kaplan:2009ag,Falkowski:2011xh}, decaying dark matter \cite{Falkowski:2009yz}, and WIMP dark matter \cite{Escudero:2016ksa, Gonzalez-Macias:2016vxy, Macias:2015cna, Batell:2017rol}. 
The first part of this paper can be categorized into the last one. 
In particular, we will focus on the dark matter mass range between $\sim 10\MEV \AND \sim 10\GEV,$ which differs from the previous studies where the mass is greater than GeVs. 
More concretely, we take an effective field theory approach based on the strategy of simplicity, 
and focus on the simplest neutrino portal operator of dimension five, 
\beq
\laq{p2}
{\f_H \cdot L \p \f \o M},
\eeq
where $\p$ ($\phi$) is a Majorana fermion
(a real scalar) carrying a hidden $Z_2$ charge, and $1\o M$ is a dimensionful coupling. Therefore the lighter one of $\p$ and $\f$ is stable.

We point out that $\p$ and $\f$ are restricted to nearly degenerate to satisfy the neutrino mass constraint from the observations of cosmic microwave background (CMB) and baryon acoustic oscillations (BAO) \cite{Ade:2015xua}, otherwise a sizable neutrino mass would be produced radiatively from the neutrino portal interaction. 
This allows the Higgs boson decay into $\p,\f$ and neutrino kinematically. Such an invisible 
decay rate will be measured in several future lepton colliders, such as the Circular Electron Positron Collider 
(CEPC), International Linear Collider (ILC), FCC-ee, and Compact Linear Collider (CLIC) \cite{CEPC, 
CEPC2,Asner:2013psa,dEnterria:2016sca,Abramowicz:2016zbo}, and thus could be a probe of the dark matter or neutrino physics.

We show that the observed dark matter abundance can be thermally explained with (co)annihilation of the WIMPs through the neutrino portal interaction. Furthermore, this dark matter, if heavier than around $2\GEV$, can be tested in the future lepton colliders.

In the second part, we study the neutrino propagation in the universe with the neutrino portal interaction. 
We show that the neutrino propagation is affected only when the invisible decay of the Higgs boson is at a rate 
fully within the sensitivity reach of the future lepton colliders.
This possibility is interesting because in the IceCube neutrino observatory \cite{Aartsen:2013jdh, Aartsen:2014gkd} the cosmic-ray neutrino event above PeVs is not yet detected especially for the Greisen-Zatsepin-Kuzmin (GZK) neutrinos \cite{Ahlers:2010fw,Gelmini:2011kg,Liu:2016brs}. Also, Glashow resonance \cite{Glashow:1960zz} is not observed. 
We point out that the absence of the high energy cosmic-ray neutrinos can be explained if the annihilation of the neutrino-(anti)neutrino into WIMPs take place before the neutrino arrives at the earth.
Namely, a cutoff for neutrino can be set from the neutrino portal interaction. 
Moreover, a neutrino mass is constrained kinematically from the mass range of the WIMPs with a given cutoff, e.g. for a cutoff of a few PeVs which could explain the non-observation of the Glashow resonance, one of the neutrino mass is within $0.01-0.2\EV$. On the other hand, for a cutoff around 10~PeV which may explain the non-detection of the GZK neutrino, one of the neutrino mass is within $0.008-0.1\EV$. Namely, a neutrino mass can be ``measured'' kinematically through the neutrino portal interaction.

A UV model is built to justify the setup and to study the experimental constraints for the heavy particles relevant for generating the higher dimensional term. 
In this model, the neutrino mass can be dominantly obtained from the neutrino portal interaction.

This paper is organized as follows. In \Sec{2} we will explain the model with several constraints and show that $\f$ or $\p$ can explain the dark matter abundance thermally.
In \Sec{3} the propagation of the cosmic-ray neutrino with the neutrino portal interaction will be discussed.
In \Sec{UVc} the UV model will be discussed. The last section is devoted to conclusions and discussion.

\section{A simple Effective Theory for WIMP}
\lac{2}

To simplify the discussion, suppose that the additional Lagrangian to that of the SM, ${\cal L}_{SM}$, has only one generation of neutrino,
\beq
\laq{lag}
\d{\cal L}= \p \ol{\s}_{\mu}\partial^\m \ol{\p} + {1\o 2}\partial^\mu \f \partial_\mu \f-{\f_H\cdot L \p \phi \o M} -{M_\p \o 2} \p \p+ h.c -  {m_\f^2\o 2} \f^2 -V(\f,\f_H), 
\eeq 
where the total Lagrangian is given by ${\cal L}={\cal L}_{SM}+\d{\cal L}$; $M_\p$ ($m_\f$) is the mass of $\p$ ($\phi$);  $V(\f, \f_H)$ is the potential of 
the scalar fields which is supposed to give a vanishing vacuum expectation value (VEV), $\vev{\f}=0$, and additional mass squared, $\vev{{\partial^2{V} \o \f^2}}=0$, to $\f$. 
We will neglect the Higgs portal term, $\l_H \f^2 \ab{\f_H}^2$, in $V(\f,\f_H)$, 
because the scalar mass is lighter than $10 \GEV$, and $\l_H$ is sufficiently small if $\l_H \lesssim {m_{\f}^2 \o v^2}$,\footnote{The neutrino portal models with an 
efficient Higgs portal interaction are studied in \cite{Escudero:2016ksa, Macias:2015cna, Gonzalez-Macias:2016vxy, Batell:2017rol}.}  where $v= 174\GEV$ is the VEV of the Higgs field.
A small portal coupling larger than the order ${1\over 16\pi^2}\({\L_{\rm c.o}\over M}\)^2$ is stable under quantum correction, where $\L_{\rm c.o}$ is the cut off scale of the model which could be smaller than $M$. The other dimension-five operators, ${(\f_H \cdot L)^2}, F_Y^{\m\n} \p 
\ol{\s}_\m \s_\n \p, \ab{\f_H}^2\p^2,\AND \f^2 \p^2$ are suppressed due to 
approximate lepton number symmetry under which $L$ and $\ol{\p}$ have 1 while others 0.\footnote{ The coefficient of these terms are stable under quantum corrections if they are greater than ${1\over 16\pi^2} {M_\p\over M^2}$. The quantum correction for $(\f_H\cdot L)^2$ will be discussed in the following.}
In particular, we suppose that a tree-level $(\f_H \cdot L)^2-$term induces a neutrino mass smaller or of the same order of the physical one. 
Notice that the tree-level $(\f_H \cdot L)^2-$term is not generated in a UV model if all the heavy particle masses and interaction preserve lepton number (see \Sec{UVc}.).
 
\subsection{Constraint from Neutrino Mass}
\begin{figure}[!t]
\begin{center}  
 \includegraphics[width=105mm]{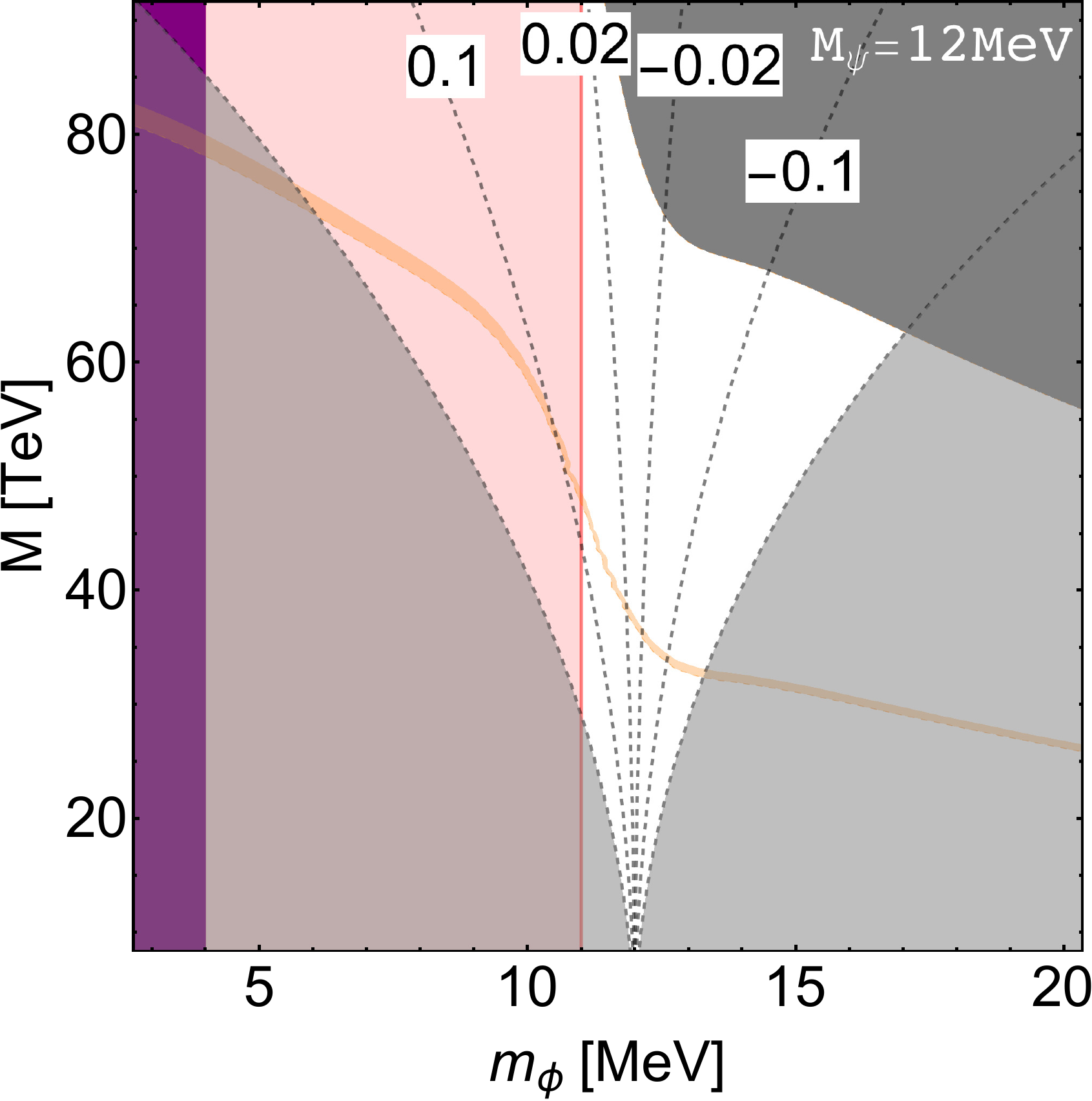}
\end{center}
\caption{ The contour plots of the radiatively generated neutrino mass [eV] with $M_\p=12~\MEV$. 
The purple region may be excluded due to the neutrino effective number. In the gray region, the universe 
is over-closed. 
On the orange band, the thermal abundance of the lighter WIMP explains the dark matter. 
The pink region may be tested in the future CMB/BAO observations. }
\label{fig:1}
\end{figure}

\lac{NM}
At the broken phase of the electroweak symmetry, one obtains an interaction
$ {v \o M} \n \p \f. $
It was pointed out that the neutrino mass is generated at the 1-loop level in this broken phase interaction\cite{Boehm:2006mi,Farzan:2009ji,Farzan:2011tz,Ma:2006km}: 
\beq
\laq{1}
m_{\n} = {1 \o 16\pi^2} { v^2 \o M^2} K M_\p+\O(({1\o16\pi^2})^2) {v^2 \o M^2}M_\p,
\eeq
where $K\equiv K\({m_\f^2 \o M_\p^2}\)$ with $K(x)=1-{x \o x-1 } \log{({x})}$ satisfying $\lim_{x\rightarrow1}K(x) = 1-x $. We have taken the renormalization scale $\m=M_\p$ so that this is an on-shell renormalization. 
Since is constrained by the CMB and BAO observations \cite{Ade:2015xua} as
\beq
\laq{nr}
m_\n \lesssim 0.2 \EV (95\% {\rm CL}).
\eeq
while is also constrained by the double beta decay experiment for an electron neutrino \cite{KamLAND-Zen:2016pfg}, the neutrino mass crucially restricts the mass range of the two WIMPs.
In the Fig.\ref{fig:1}, the contour plot of the generated neutrino mass and the constraint on it (gray shaded region) are represented in $m_\f-M$ plane with $M_\p=12\MEV$. 
One sees that $m_\f$ is restricted to be around $M_\p$, 
and the smaller the $M$, the smaller the difference $\ab{m_\f-M_\p}$. 
Since one of $\p$ and $\f$ is stable, $M$ has an upper bound for sufficient (co)annihilation of $\f$ or $\p$ not to over-close the universe. Thus, $\f$ and $\p$ are constrained to be nearly degenerated,
\beq
m_\f \simeq M_\p.
\eeq
Notice that this constraint disappears when $\f$ is to be replaced by a complex scalar field with only a bilinear mass term because lepton number symmetry recovers. 
However, let me pursue on the simple real scalar case with $m_\f\simeq M_\p$, but the following discussion will be qualitatively the same in a specific parameter region with complex extension of the scalar field.

\subsection{Heavy Boson Decays in Colliders}

Since $m_\f +M_\p \lesssim 20\GEV$ under our consideration, the anomalous decays of the Higgs, $W$- and $Z$- bosons into $\p,\f$ and a lepton are possible.
Thus, in colliders this scenario is constrained and tested.
In particular, the Higgs boson invisible decay is represented as
\beq
\laq{Hdecay}
H \rightarrow \p +\f+\n~(\ol{\n}).
\eeq
The decay width of the process is obtained as 
\beq
\G_{H \rightarrow inv}\simeq {1\o 1536 \pi^3}{m_H^3 \o M^2},
\eeq
where $m_H$ is the Higgs boson mass and the decay products are approximated to be massless. 
Given the total decay width of the Higgs boson $\simeq 4\MEV$, the branching ratio of this process are estimated as
\beq
\laq{br}
{Br_{H\rightarrow inv}} \simeq 0.01\% \({10 \TEV \o M}\)^2,
\eeq
 where the bound from the LHC is ${Br_{H\rightarrow inv}}<25\% (95\% {\rm CL})$ \cite{Aad:2015pla,Khachatryan:2016whc}. 
 
On the other hand, the decay rate of $W$-boson to a charged lepton and missing energy ($Z$-boson to missing energy) can be estimated as $\G_{W^{-}\rightarrow l^-+{\rm missing}}\simeq \G^{tree}_{W^{-}\rightarrow l^- +\ol{\nu}_l} (1-{1\o 24\pi^2}({v\o M})^2)$ ($\G_{Z\rightarrow {\rm missing}}\simeq \G^{tree}_{Z\rightarrow \ol{\n}+ {\n} }(3-{1\o 12\pi^2}({v\o M})^2)$) at the leading order of the anomalous decay\footnote{The processes with virtual $\f,\p$ emission and absorption are also included in the decay width.}, where $\G^{tree}_{W^{-}\rightarrow l^- +\ol{\nu}_l}$ ($\G^{tree}_{Z\rightarrow {\n}_l +\ol{\nu}_l}$) is the decay rate of the subscript at the tree-level in the SM.
The branching ratio of $W$-boson to lepton + missing ($Z$-boson to missing) differs from the SM one by $1\times 10^{-6}\% \({10 \TEV \o M}\)^2$ ($8 \times 10^{-7} \%\({10 \TEV \o M}\)^2$). The corresponding LEP bound is given as $0.1\%$ ($0.06\%$)\cite{Olive:2016xmw}.

One finds that $M$ can be as small as $\O(100)\GEV$ to be consistent with the current experiments.  To be conservative, let us set a bound\footnote{For $M\lesssim 1\TEV$, one may care for the constraint for a heavy field in some UV models.  The constraint in a UV model, which will be discussed \Sec{UVc}, is represented by a lower bound \eq{UVM} similar to \eq{Mband}.}
\eqn{ \laq{Mband}M\gtrsim 400\GEV.} 
This is represented as the horizontal black band in Fig.\ref{fig:2}.

On the other hand, the Higgs invisible decay with \eqn{\laq{CEPC}M\lesssim 5\TEV} can be tested in the future lepton colliders, such as the CEPC, ILC, FCC-ee, and CLIC, where the branching ratio of the invisible decay is planned to be measured at a precision around $0.1\%$ (the purple shaded region in Fig.\ref{fig:2}.)\cite{CEPC, 
CEPC2,Asner:2013psa,dEnterria:2016sca,Abramowicz:2016zbo}.

\subsection{Thermal Relic Abundance of WIMP}
\begin{figure}[!t]
\begin{center}  
 \includegraphics[width=105mm]{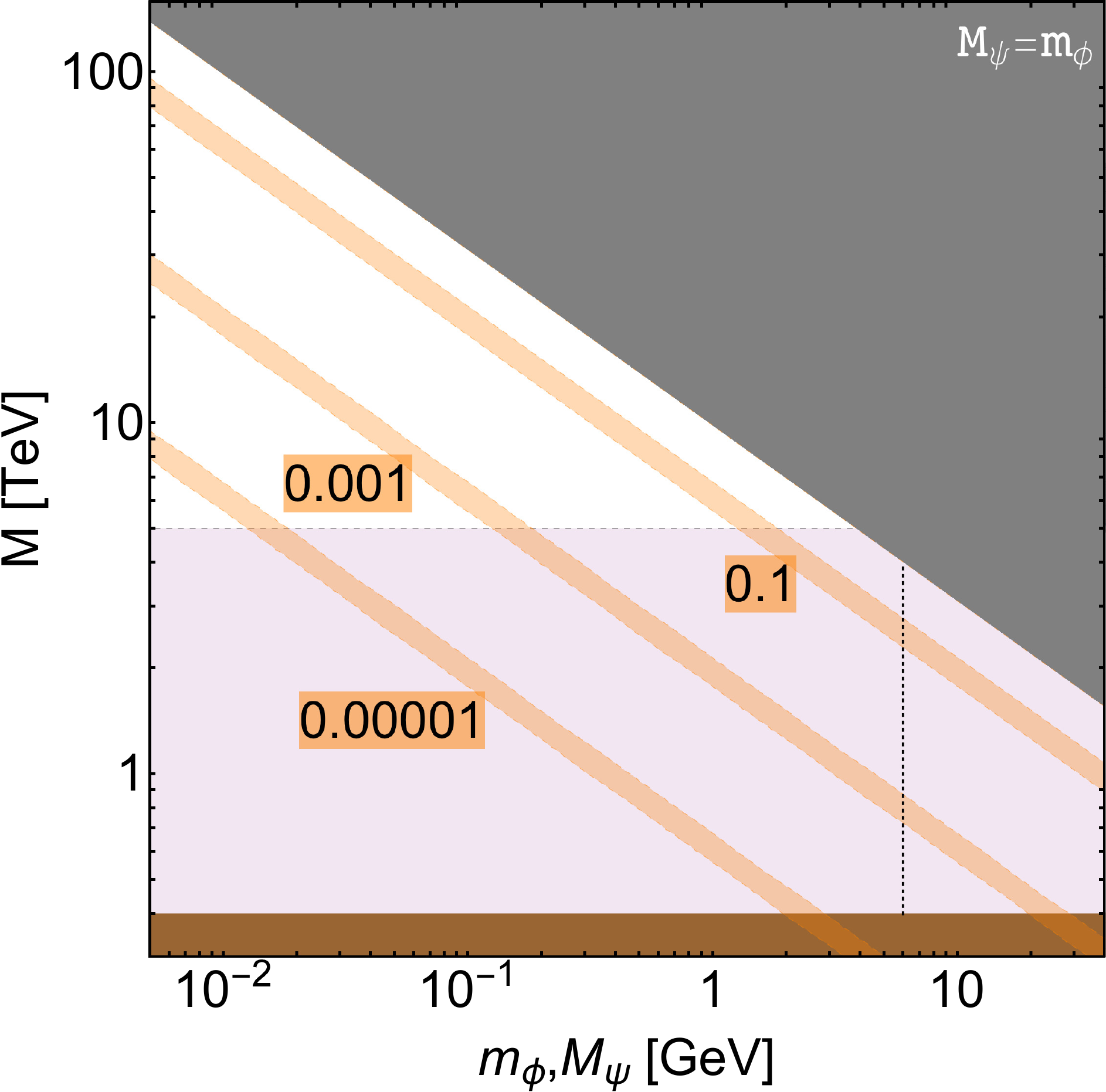}
\end{center}
\caption{ The contour plots of the WIMP relic abundance with $M_\p\simeq m_\f$. 
The brown region is excluded by heavy boson decays in the colliders. 
The black dotted line denotes $M_\p\simeq m_\f=6\GEV$. The light pink region is testable in the future lepton colliders by measuring the Higgs boson invisible decay rate.}
\label{fig:2}
\end{figure}

Now let us discuss the thermal relic abundance for the lighter of $\f$ or $\p$. 
The lighter one annihilates into (anti-)neutrinos through t(u)-channel,
\beq
\f + \f \rightarrow  \n +\ol{\n}~~ (m_\f<M_\p),\ebq
\p+\p \rightarrow \n+\n,\ol{\n}+\ol{\n}, \n +\ol{\n}~~ (M_\p<m_\f).
\eeq
In the first low, one does not have $\f+\f\rightarrow \n +\n \OR \ol{\n}+\ol{\n}$, because
the corresponding effective vertex by integrating out $\p$ vanishes with the equation of motion for external neutrinos.
The total annihilation cross sections times the relative velocity at the tree-level are given as,
 \beq\laq{anDM}
v_{\rm rel}\s_{\p\p}(s) \simeq {1\o 16 \pi} \({{v^2 \o M^2}}\)^2 {M_\p^2 \o (m_\f^2+M_\p^2)^2}\( 1 +\O\({s\o 4m_\f^2}\)\),\ebq 
\non
 \AND v_{\rm rel} \s_{\f\f}(s) \simeq {1 \o 2 \pi} \({{v^2 \o M^2}}\)^2 {M_\p^2 \o (m_\f^2+M_\p^2)^2}\(1 +\O\({s\o 4M_\p^2}\)\),\eeq 
for annihilations of $\p\p$ and $\f\f$, respectively.
The $\O(s)$-terms are calculated by FeynRules and FormCalc 
\cite{Alloul:2013bka,Hahn:1998yk}. FeynRules and FormCalc are also used to confirm all of the amplitude calculations in this paper.
The dark matter abundance is estimated as 
\beq
\laq{DMab}
\Omega_{\f,\p} h^2=0.1 \({ 4 \times 10^{-26}  {\rm cm^3/s}  \o \vev{\s_{eff} v} } \)\( x_f \sqrt{g_{*}} \o 5 g_{*s} \) ,\eeq 
where $\vev{\s_{eff} v}$ is the thermal averaged annihilation cross section given by $$\vev{\s_{eff} v} = {\sum_{i}{ g_i^2\int_{2m_i^2}{ds \sqrt{s} K_1(\sqrt{s}/T) \(s/4-m^2_i\) \s_{i}(s)}}\o 2T(\sum_{i}{g_i m_i^2K_2(m_i/T))})^2}$$ ($K_j(x)$ is the modified Bessel function of the $j-$th kind)
  \cite{Belanger:2001fz}, where the coannihillation effect is included; $g_* (g_{*s})$ is the degree of freedom for 
the energy (entropy) density of the radiation which is typically around $10-100$ for $\O(10)\MEV<m_{DM}\lesssim 10\GEV$; $x_f={  m_{DM}\o T_f}$ is the freeze-out temperature in the unit of $m_{DM}=\min{(m_\f, M_\p)}$ which is around $15-20$; $h=0.678$. 
 The region satisfying $\Omega_{\f,\p}h^2\simeq 0.1$ is represented by the orange band in Fig.\ref{fig:1}. 
In Fig. \ref{fig:2}, the contours of $\Omega_{\f,\p}h^2$ are shown (orange bands) at the limit $M_\p=m_\f$. The width of the orange bands denotes the ambiguity of our calculation.\footnote{  The width of a band in the figure is obtained by using the largest and smallest $x_f\sqrt{g_*}\o \sqrt{g_{*s}}$ on the band. }
The gray regions in both figures denote the over-closure of the universe, $\Omega_{\f,\p}\gtrsim 1$. 
In this figure, one finds that the thermal dark matter can be tested in the future lepton colliders with mass
\beq
m_\f \simeq M_\p\gtrsim 2\GEV.
\eeq 
If the lighter of $\f$ or $\p$ is part of the dark matter, the testable mass range increases.

In particular, the mass greater than $6\GEV$ is now testing in Xenon1T, LUX, and PandaX \cite{Aalbers:2016jon,
Akerib:2015cja,
Aprile:2015uzo} (This boundary is represented as the black dotted line in Fig.\ref{fig:2}.), and it is interesting that we can have a cross-check if the dark matter is detected in the direct-detection experiments.

 \subsection{$N_{eff}$ and BBN}

\lac{neutrino}
The mass of $\f$ or $\p$ should be larger than $\MEV$s, otherwise the created neutrinos from the annihilation of them or themselves will change $N_{eff}$ by $\O(1)$ and could spoil the BBN \cite{Kolb:1986nf,Serpico:2004nm,Boehm:2013jpa,Nollett:2014lwa}. 
According to \cite{Nollett:2014lwa}, 
\beq
\non
m_\f > 5\MEV, M_\p > 7\MEV~ (\ab{m_\f-M_\p} \gtrsim {\rm MeVs} )\ebq 
\laq{massr}
m_\f, M_\p > 9\MEV~ (m_\f\simeq M_\p)
\eeq
is obtained from the bound for $N_{eff}$ \cite{Ade:2015xua}.

On the other hand, the viable region with \beq m_\f \OR M_\p \lesssim11\MEV \eeq has a slightly larger $N_{eff}$ and may be tested by several future CMB observations such as the PIXIE and CMB-S4 experiments, as well as the BAO observation \cite{Kogut:2011xw,Abazajian:2016yjj,Baumann:2017lmt}.


\section{Propagation of Cosmic-Ray Neutrino with Neutrino Portal Interaction}
\lac{3}
\lac{GZK}

Now we focus on the region where the lighter WIMP composes a fraction of the dark matter, $\Omega_{\f,\p}h^2< 0.1$, i.e. the region with sufficiently strong neutrino portal interaction.
The observed dark matter abundance can be explained with other dark matter components: a WIMP with neutrino portal interaction of different generation (see \Sec{5} and footnote \ref{ft:1}), a superpartner\footnote{There are several typical lightest superpartners (LSPs) which might be the dominant dark matter, depending on SUSY breaking scenarios: gravitino LSP in gauge mediation \cite{Giudice:1998bp}, bino-like LSP with SUSY breaking in a gauge unified manner, wino-like LSP in anomaly mediation and simple SUSY breaking scenarios based on the anomaly mediation \cite{ANSBLS,
ANSBM,
Pomarol:1999ie,
Chacko:1999am,
Ibe:2006de,
Ibe:2011aa,
ArkaniHamed:2012gw,
Yin:2016shg,
NGH}, $N=2$ superpartners in $N=2$ partial breaking \cite{Nojiri:2007jm,Belanger:2009wf,
Polonsky:2000zt,
Benakli:2009mk,
Chun:2009zx,
DeSimone:2010tf,
Heikinheimo:2011fk,
Benakli:2012cy,
Dudas:2013gga,
Benakli:2014cia,
Goodsell:2015ura,
Shimizu:2015ara,
Yin:2016pkz}, etc.  }, an inflaton \cite{Kofman:1994rk,Kofman:1997yn,Mukaida:2014kpa, Lerner:2009xg,Okada:2010jd,Khoze:2013uia,Bastero-Gil:2015lga,Daido:2017wwb, Nakayama:2010kt, Chen:2017rpn}, etc.
This region is interesting because it would affect the propagation of the neutrino in the universe.

Although more statistics are needed, up to now no cosmic-ray neutrino event above several PeVs is detected in the IceCube experiment \cite{Aartsen:2013jdh, Aartsen:2014gkd},
and the Glashow resonance around $6$~PeV is also not observed \cite{Glashow:1960zz}.
Despite several detections of cosmic-ray events of other kind particles up to $\sim 10^{2}$~EeV, this fact implies that there may be a special cutoff for the cosmic-ray neutrino.
In particular, if some of the observed cosmic-rays around $10^{2}$~EeV are protons, cosmic-ray neutrinos of $
\O$(EeV) should also be obserbed. Protons of energy larger than $\O(10^2)$~EeV interacts with a CMB-photon 
and produces pions via $\D$-resonance, and hence loses energy before the cosmic-ray neutrino reaches the 
earth. This scattering sets a GZK cutoff at energy $\O(10^2)$~EeV for protons \cite{Zatsepin:1966jv,Greisen:1966jv} which explains the observed cutoff for high energy cosmic-ray events. 
In the GZK cutoff scenario, GZK neutrinos of energy $\O$(EeV) are produced from the decay of these pions\cite{Beresinsky:1969qj} and should be detected at $\O(0.1)-\O(10)$ events/year in the IceCube neutrino observatory \cite{Ahlers:2010fw,Gelmini:2011kg,Liu:2016brs}.

The non-detection of such energetic neutrino events can be explained from a viewpoint of particle physics.\footnote{There are also astronomical explanations for the absence of neutrino events above several PeVs. For example, if the neutrino is originated from the galaxy clusters or starburst galaxies the non-observation of the Glashow resonance can be accounted for \cite{Murase:2016gly}. If the $\O(10^2)$~EeV cosmic-rays observed are composed of heavy nuclei such as iron, the absence of the GZK neutrino event can be explained \cite{Iron}.  }
Thanks to the neutrino portal interaction, annihilation between the cosmic-ray and the cosmic background neutrinos,  \eqn{\laq{nnan} \n/\ol{\n} + \n ({\rm C\n B}) \rightarrow \p+\p , \f+\f,} is enhanced with sufficiently small $M$ so that before the neutrino reaches the earth it turns into the WIMPs.\footnote{The explanation of the neutrino events especially for absorption lines in the observed neutrino flux at IceCube, in the light of the interaction between a cosmic-ray neutrino and a cosmic background neutrino, was discussed in several recent studies.  \cite{Ioka:2014kca, 
Cherry:2014xra, 
Ng:2014pca,
Ibe:2014pja,
Blum:2014ewa,
Araki:2014ona,
DiFranzo:2015qea,
Araki:2015mya,
Shoemaker:2015qul}. }
Namely, we propose that the neutrino portal interaction can set a cutoff for cosmic-ray neutrinos.

To set a cutoff, there are two conditions. 
First, the annihilation channel should be turned on at $E_\n > E_{\n}^{cutoff}$, and hence the center of mass energy of the neutrino-(anti)neutrino system, $E_{cm}$, should become greater than the threshold, $2M_\p \OR 2m_\f$, at $E_{\n}^{cutoff}$, as
\beq
\laq{Ecm}
 {E_{cm}\o 2} \equiv {1\o \sqrt{2}}\sqrt{m_\n^2+E_\n E_{\rm C\n B}(1-\cos{\h})} \gtrsim M_\p \OR m_\f
 \eeq
where $E_{\n}^{cutoff}$ is defined at the equality
\beq
\laq{mass}
M_\p \OR m_\f \sim \({E_\n^{cutoff} \o 6{\rm ~PeV}} {E_{\rm C\n B}\o 0.2\EV}\)^{1\o 2} 35 \MEV.
\eeq
Here $E_{\rm C\n B} \simeq \max{[T_\n,m_\n]}$ is the typical energy of cosmic background neutrinos with temperature $T_\n \simeq 2\times 10^{-4} \EV$, and ${\h}$ is the angle between the momenta of two neutrinos. 

Secondly, the mean free path, $d({E_\n})$, imposed by the annihilation should be shorter than the distance to the neutrino source. To discuss this, let us neglect for simplicity the neutrino oscillation.\footnote{\label{ft:1} Given the neutrino oscillation, all kinds of the neutrinos share the strongest neutrino interaction and the mean free path for each neutrino should be \Eq{GZK} times a factor $\sim \O(1)$. Thus, in the multi-generation extension of the neutrino portal interaction, one does not need all the interactions to be strong to set the cutoff for different kinds of neutrinos, and this allows one of the WIMPs becomes the dominant dark matter. }
Following \cite{Ibe:2014pja}, one obtains the mean free path of a neutrino given by
\beq
\laq{GZK}
d(E_\n) \simeq  \int{{d^3\vec{p}\o (2\pi)^3}\s_{\n\n} (E_{cm}(\vec{p},E_\n)) f_{C \n B}(\vec{p})  }.
\eeq
Here $f_{C\nu B}(\vec{p})=2 (e^{\ab{\vec{p}}/T_\n} + 1)^{-1}$ is the neutrino distribution function for the cosmic background neutrinos, and $\s_{\n\n} (E_{cm} )$ is the helicity averaged neutrino-(anti)neutrino annihilation cross section. 
This annihilation cross section with $m_\f\simeq M_\p$ is approximated as
\beq
\laq{nn}
\s_{\n\n}(E_{cm})\simeq{v^4 \o M^4}\frac{\sqrt{\({{E_{cm}\o 2}}\)^2-{m_\f}^2}+ {E_{cm}} \log \left(\frac{\sqrt{\({E_{cm} \o 2}\)^2-{m_\f}^2}+{E_{cm}\o 2}}{{m_\f}}\right)}{16 \pi E_{cm}^3}.
\eeq
Then the neutrino flux from the source at $L$ distant place is weakened by a factor of
\beq
\laq{sup}
\k(E_\n)=e^{-{L \o d(E_\n)}}.
\eeq 
Here we have neglected the effect of the redshift for $E_\n$ due to the expansion of the 
universe, which would reduce the observed $E_\n$ in the IceCube by $\O(10)\%$ with $L\sim \O(1)$Gpc.

\begin{figure}[!t]
   \begin{center}  
   \includegraphics[width=100mm]{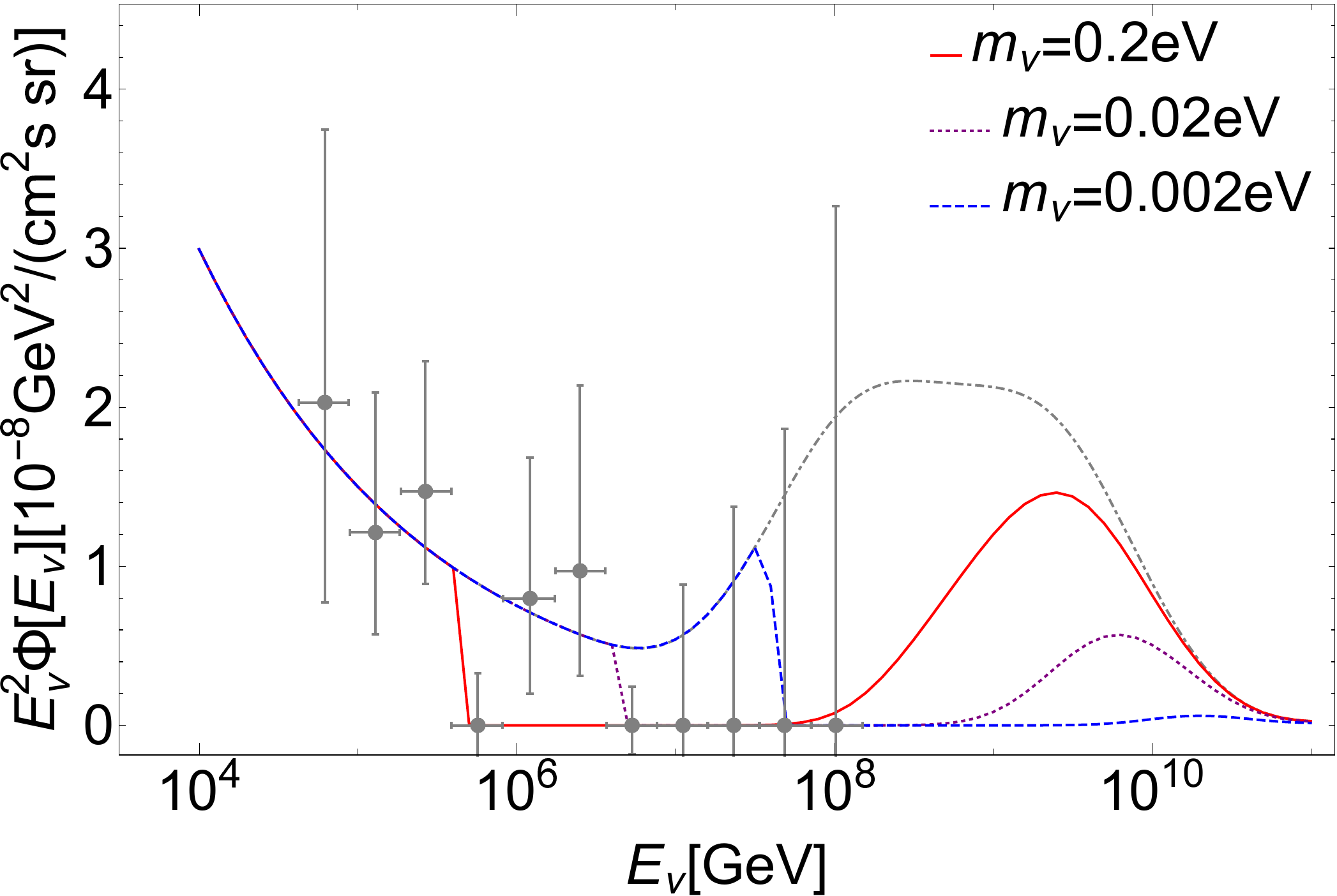}
   \end{center}
\caption{The predicted neutrino flux with several $m_\n$. $L=300$~Mpc, $M_\p=m_\f=9\MEV$ and $M=450\GEV$ are fixed. The red solid, purple dotted, and blue dashed lines represent the flux with $m_\n=$ $0.2\EV$, $0.02\EV$, and $0.002\EV$, respectively. The gray points represent the IceCube observation arranged from \cite{Aartsen:2014gkd} while the gray dot-dashed line represents the flux distribution before the annihilation \Eq{fit}. }
\label{fig:3}
\end{figure}

For instance, the predicted neutrino flux is represented in Fig.\ref{fig:3} by assuming a neutrino flux before the annihilation as \beq \laq{fit}
\F(E_\n) E_\n^2=1.5\times10^{-8}\({E_\n\o 10^5 \GEV}\)^{-0.3}+V_{GZK}(E_\n).
\eeq 
 The first term is the best-fit power law in \cite{Aartsen:2014gkd} while the second term represents a toy GZK neutrino flux, \beq
 \non
 10^{8} V_{GZK}(E_\n) = \(e^{\({3 \o 8}\cos{\(\pi{\log_{10}{(E_\n/{\rm GeV})}-8.75 \o 2.5}\)}-{1\o 
24}\cos{\(3\pi{ \log_{10}{(E_\n/{\rm GeV})}-8.75 \o 2.5}\)}\)}-e^{-{1\o 3}}\) \eeq
for $E_\n>10^{6.25}\GEV$, otherwise 0. (The realistic ones for GZK neutrino are given in \cite{Ahlers:2010fw,Gelmini:2011kg,Liu:2016brs}.) 
The neutrino flux in our scenario is approximated as 
\beq
\laq{pfl}
\k(E_\n) \F(E_\n) E_\n^2.
\eeq  

In Fig.\ref{fig:3}, one finds the neutrino flux does get a cutoff or an absorption band through the $t$-channel annihilation. Notice that the cut-off is less efficient in a model where there is a significant $s$-channel annihilation/scattering process. In fact, the $s$-channel process itself does not contribute like a ``cutoff" but an absorption line due to the quick decrease of the cross section when the center of mass energy exceeds the threshold. 
Moreover, the scattering process between neutrino and (anti)neutrino is at tree-level if $s$-channel process exists. This is constrained by the CMB observation \cite{Cyr-Racine:2013jua} and the efficiency of the cutoff is bounded. In our case the scattering process is 1-loop suppressed and this bound is much looser than the heavy boson decay.

\subsubsection*{Relation with Heavy Boson Decay }

\begin{figure}[!t]
\begin{center}  
   \includegraphics[width=72.5mm]{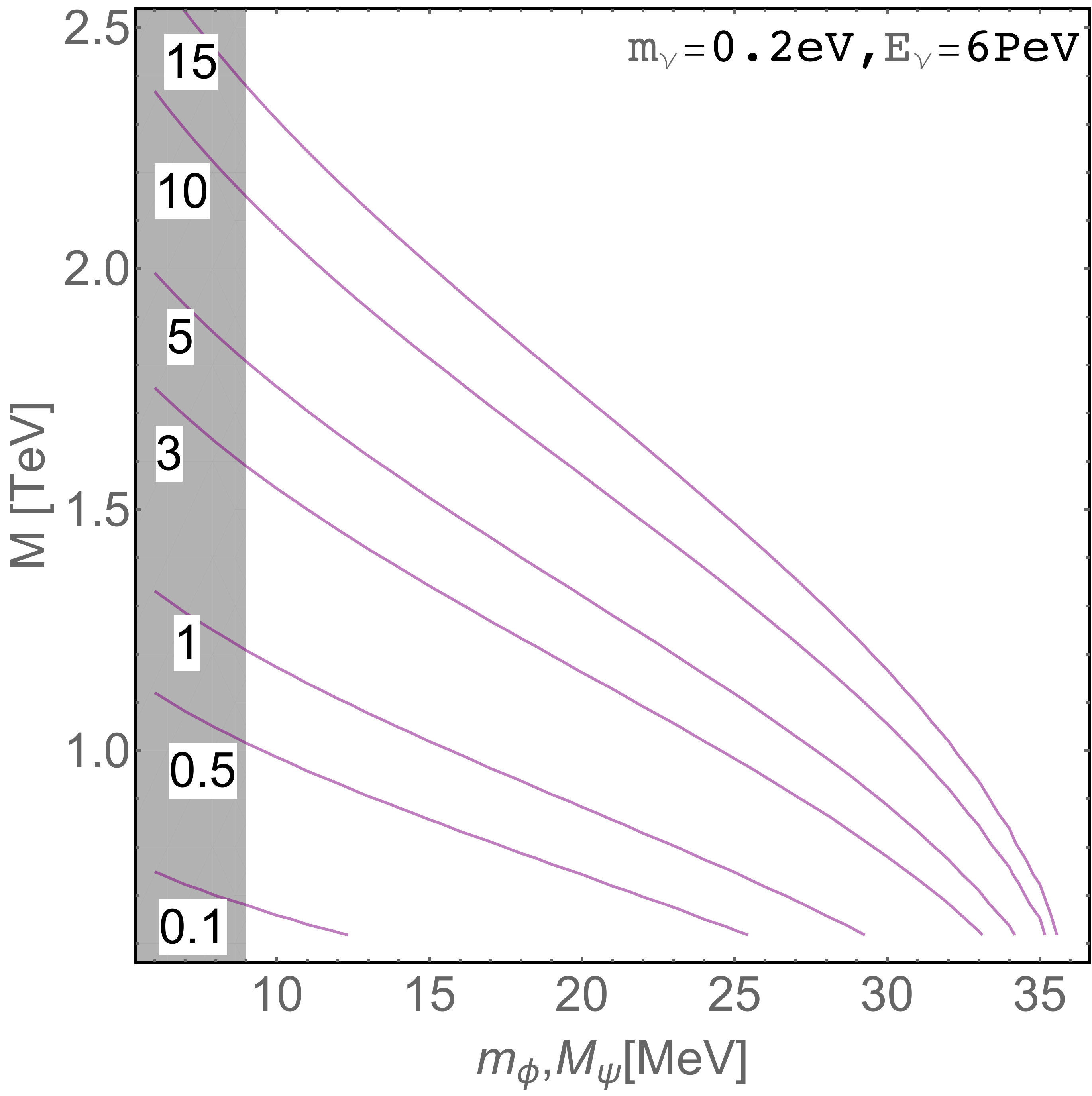}
   \includegraphics[width=75.5mm]{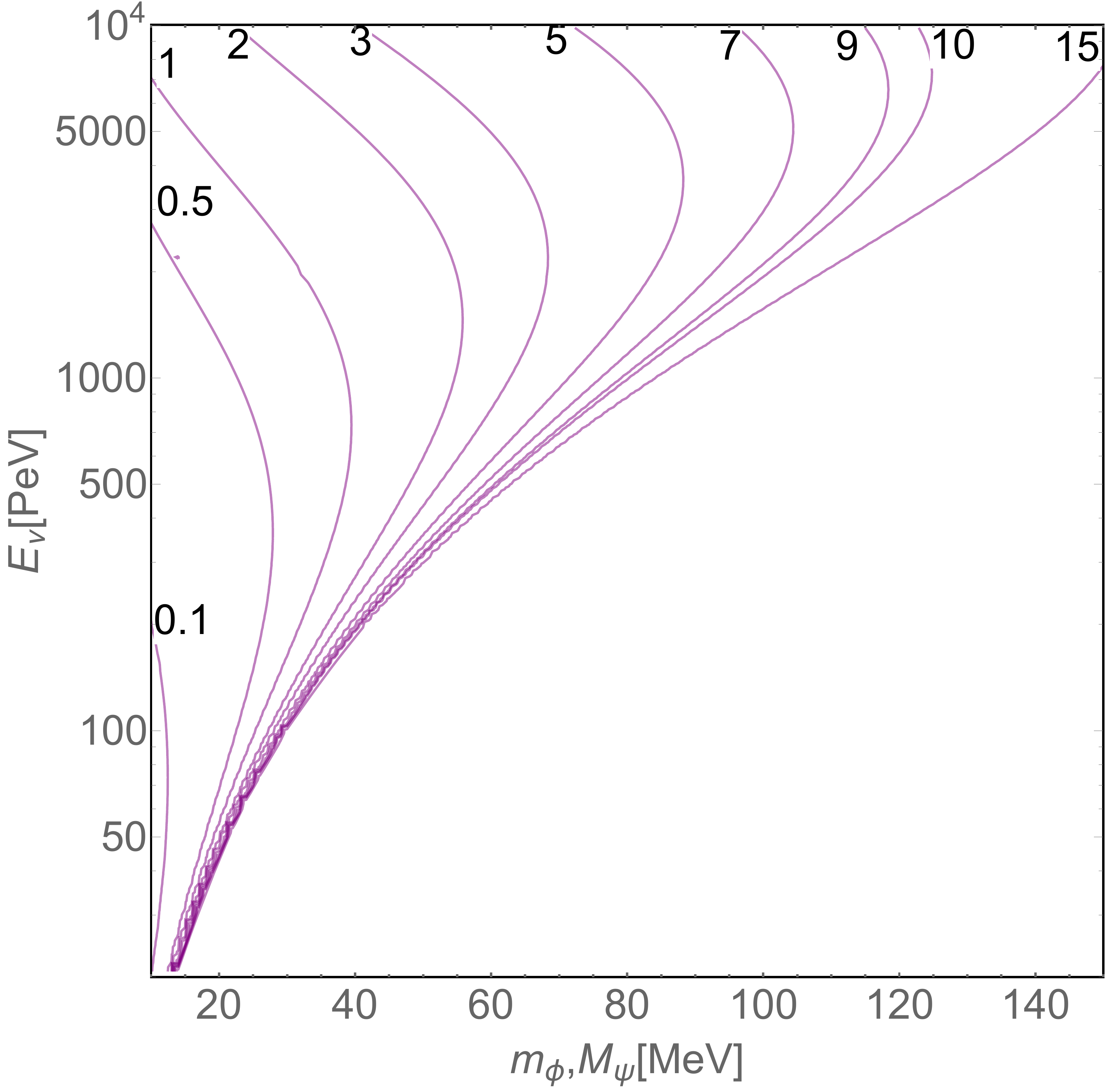}
\end{center}
\caption{ The contour plot of the mean free path $d(E_\n)$ [Gpc] for neutrino with $M_\p \simeq m_\f$. In the left panel, $E_\n=6$~PeV and $m_\n=0.2\EV$ is fixed. 
The vertical shaded region represents the constrained region from the neutrino effective number. 
In the right panel $M=630\GEV \AND m_{\n}=0.01\EV$ are fixed. }
\label{fig:GZKcp}
\end{figure}

The contour plot of $d(E_\n)$ is represented in Fig.\ref{fig:GZKcp}. 
From the left panel, one finds that the neutrino flux originating from a place with
\beq L > \O(10){\rm ~Mpc} \eeq can be affected with the neutrino portal interaction. 
 
As in the left panel of Fig. \ref{fig:GZKcp}, we have checked that to obtain $d(E_\n)$ 
smaller than the scale of particle horizon size $\sim 10$~Gpc, i.e. when the neutrino propagation in the universe could be affected, $M$ should be smaller than $\sim 2\TEV$. 
From the right panel, where $M$ is at around the lower bound \eq{Mband},
one reads the upper bound of $M_\p\simeq m_\f$ to be around $ \O(100)\MEV$. 
This upper bound becomes smaller with larger $M$ due to the scaling of the cross section. 
Hence one obtains the parameter range where the neutrino propagation in the universe is affected,
\beq 
M\simeq 0.4 - 2 \TEV \AND \laq{massr2} 
m_\f \simeq M_\p \sim 9-\O(100) \MEV.\eeq
Since the upper bound of $M$ satisfies \eq{CEPC}, the following is predicted:
if the high energy neutrino flux in the IceCube is affected by the neutrino portal interaction, the Higgs invisible decay is fully within the reach of the future lepton colliders. 
Let us emphasize again that $M$ required here is much smaller than the one for \Eq{DMab}, and we can not provide dominant dark matter whose interaction affects the cosmic-ray neutrinos. 
However, to explain the neutrino oscillation an extension with several flavors of $\p$ or $\f$ is needed. (See conclusions and discussion.) In this case, 
some of the flavors can be the dominant dark matter while some can affect the spectra of cosmic-ray neutrinos. We note that in this case the dark matter should be lighter than the particles relevant to the cutoff, and thus the dark matter mass is lighter than $\O(100)\MEV.$ 

\subsubsection*{Measuring Physical Neutrino Mass Range}

The neutrino flux carries the information of the annihilation during its propagation. In particular, as in a collider, 
one can measure $E_{\rm C\nu B}$ kinematically once the cutoff and the masses of $\f$ and $\p$ are given in 
someway. This implies a mass scale can be obtained for one of the neutrinos if its mass is greater than $T_\n$.

Even just given the mass range of $\f, \p$, one can predict the neutrino mass range.
Since $E_{\rm C\n B}E_{\n}^{cutoff} \sim m_\f^2 \simeq M_\p^2 $, with the cutoff scale fixed, one obtains
\beq
E_{\rm C\n B} \propto m_\f^2 \simeq M_\p^2
\eeq
which implies $E_{\rm C\n B}$ ($m_\f \simeq M_\p$) has a lower bound corresponding to \Eq{massr} ($E_{\rm C\n B}\gtrsim T_\n$). 
If $E_{\rm C\n B}>T_\n$ at the lower bound of \Eq{massr}, the lower bound of one of neutrino masses is predicted.

For instance, for a cutoff at $6$~PeV one obtains a lower bound of neutrino mass $m_\n\gtrsim1.4 \times10^{-2}\EV$. 
With a cutoff $\lesssim 6$~PeV, which may explain the non-observation of the Glashow resonance, the neutrino mass lower bound becomes greater. 
 Thus the neutrino mass range is predicted as 
\beq 
m_\n \simeq 0.01-0.2\EV~ ({\rm for} ~E_{\n}^{cutoff} \lesssim 6 {\rm ~PeV}).
\eeq
The neutrino flux around the lower limit is illustrated by the purple dotted line in Fig.\ref{fig:3}.
This mass range covers the atmospheric neutrino scale $0.05\EV$.

If the GZK neutrino source is originated from $\simeq \O(1)$Gpc away from the earth, $m_\f \AND M_\p$ should be smaller than $\O(100)\MEV$. This can be found in the right panel of Fig.\ref{fig:GZKcp} because for any $E_\n m_\n$ this is almost satisfied. 
Then, for a cutoff around $10$~PeV, which may explain the non-detection of the GZK neutrino (See the blue dashed line in Fig.\ref{fig:3}), the neutrino mass range can be estimated as 
\beq
0.008\EV \lesssim m_\n \lesssim 0.1 \EV~ ({\rm for} ~E_\n^{cutoff}\simeq10{\rm ~PeV}).
\eeq
The lower bound, where the annihilation is most efficient, is close to the solar neutrino scale of $0.009\EV$.

\section{A UV model}
\lac{UVc}
In the previous sections, we have studied a dimension 5 operator with two additional gauge singlets. 
It is questioned that whether there is a UV model, and if is, whether constraints for heavy particles in the UV model restrict our scenario especially for $M\lesssim \TEV$.

To suppress $(\f_H \cdot L)^2$-term in order to satisfy the constraint \eq{nr} at the tree-level, 
the UV model should also have an approximate lepton number conservation.
One of such UV models is given by
\beq
\laq{nUV}
{\cal L}=- y\f_H L N-\tl{M}S N- {\tl{M} \o 2f} \f S \p
-{M_N\o 2} N N-{M_\p\o 2 } \p \p + h.c. -{m_\f^2\o 2} \f^2  -V(\f,\f_H)
\eeq
where $S$ and $N$ are gauge singlet Weyl fermions with lepton number 1 and -1, respectively, and we have omitted the kinetic terms.
For later convenience, we introduce a Yukawa coupling $y$, the decay constant $f$, the order parameter $\tl{M}$, and the mass parameter $M_N$ satisfying
$M_N, M_\p \ll \tl{M}$ due to the approximate lepton number symmetry.
We have forbidden $M_S SS$ term at the tree-level by imposing $Z_4$ symmetry 
under which $S,N, \psi$ and the spurion $\tl{M}$ are charged by $1/4, 1/2, 1/2 \AND 1/4 $, respectively.  
Leptons can carry ${1/2}$ so that Yukawa couplings are allowed. 
Thus this symmetry is identified to be spontaneously broken down due to the VEV  $\tl{M}$ of some scalar field.

By making a shift of $S\rightarrow S-{y\o \tl{M}}\f_H L$, one finds that the neutrino portal interaction appears as 
\beq
{\cal L}\rightarrow {1\over M} \f_H \cdot L \f  \p
-\tl{M}S N- {\tl{M} \o 2f} \f S \p-{M_N\o 2} N N-{M_\p\o 2 } \p \p + h.c. -{m_\f^2\o 2} \f^2  -V(\f,\f_H),
\eeq
with
\beq
\laq{M}
{1\o M}=-{ y \o 2f}.
\eeq
The neutrino portal term (1st term) is decoupled from the heavy fields, $S\AND N$.
 Moreover, $(\f_H \cdot L)^2$ does not appear by integrating out the heavy fields up to 1loop level, 
because $\f_H \cdot L$ does not directly couple to the heavy field. 
This fact can be also checked by integrating the heavy fields out in terms of \Eq{nUV} after diagonalizing the fermion mass matrix.
Interestingly, with this UV model, the neutrino mass is purely generated radiatively through the neutrino portal interaction.
The Higgs portal term, $\ab{\f_H}^2 \f^2$, could be suppressed if $\f$ is a pseudo-Nambu-Goldstone boson with breaking scale $f$ (See discussion for a concrete non-linear sigma model.).\footnote{ At the tree-level, the portal coupling is of order $ \sim m_\f^2/f^2$ since a Nambu-Goldstone boson for an exact global symmetry does not have potential and $m_\f$ could be the size of the explicit breaking.} 

There are several constraints for $N$, because it behaves as a right-handed neutrino with Yukawa coupling $y$ and the mass $\tl{M}$ \cite{deGouvea:2015euy, Bertoni:2014mva}.
If we adopt the constraint in
\cite{deGouvea:2015euy} for a heavy right-handed neutrino, which dominantly mixes with $\tau$ neutrino,  
$\ab{y v \o \tl{M}}\lesssim 0.1$ is required.  On the other hand, this effective theory should have ${\tl{M} \over 2f} \lesssim \sqrt{4\pi}$ from the viewpoint of perturbative unitarity, and  ${\tl{M}\over f} \lesssim 2\pi$ when $\f$ is identified as a pion-like field, which, respectively, turn out to be
\beq
\laq{UVM}
M\gtrsim {v\o 0.1 \times \sqrt{4\pi}}\sim 500\GEV \AND {v\o 0.1 \times 2\pi} \sim 300\GEV.
\eeq

\section{Conclusions and Discussion}
\lac{5}

In this paper, a simplest neutrino portal interaction for WIMPs was investigated, especially for the lightest WIMP 
mass in the range of $ \O(10)\MEV -\O(10)\GEV$, where is not yet severely constrained by direct detections. Neutrino portal interaction is interesting because it can affect not only collider physics for the Higgs boson but also neutrino physics.

We pointed out that the constraint for radiatively generated neutrino mass seriously restricts the parameter 
space so that the two WIMPs are nearly degenerate. Due to this restriction, the Higgs boson can decay into the WIMPs plus a neutrino and this invisible decay can be searched for in the future lepton colliders, CEPC, ILC, FCC-ee and CLIC. 

We showed that the neutrino portal interaction can successfully (co)annihilates the lightest WIMP and the WIMP relic abundance can explain the observed one for dark matter. Such a neutrino portal dark matter is tested in the future lepton colliders for the mass $\gtrsim 2 \GEV$.

When the WIMP explains a small fraction of the dark matter abundance, the neutrino propagation in the universe can be significantly affected. We pointed out this region can be fully tested in the future lepton colliders. In particular, this region can set a cutoff for cosmic-ray neutrino and can explain the non-detection of the GZK neutrino event or the non-observation of the Glashow resonance in the IceCube.
Moreover, a neutrino mass can be ``measured" kinematically from the scale of the cutoff and a WIMP mass.

Using a UV model, we have justified our set up and showed that a neutrino mass can be dominantly generated from the neutrino portal interaction.
\\

Since there are generations in the SM, it is natural to make an extension of the neutrino portal interaction \eq{p2} to that with 3 generation cases, e.g.
$ \f_H\cdot L^i {Y_{ij} M^{-1}} \p^j \f ( \f_H\cdot L^i {Y_{ij} M^{-1}} \p \f^j)$,
where $i,j$ denotes the generation and $Y_{ij}$ is the dimensionless coupling in the mass basis of 
$\p^j$ ($\f^j$). 
The neutrino mass matrix is generated with $m_{\n ij} \simeq  {\sum_k{Y_{ik} M_{\p k} Y_{kj} K_{k}} 
\o 16\pi^2} { v^2 \o M^2}$ ($ {\sum_k{Y_{ik}Y_{kj} K_{k}} M_\p \o 16\pi^2} { v^2 \o M^2}$) where 
$K_j$ is $K$ in \Eq{1} but with $M_\p$ ($m_\f$) to be replaced by the mass of $\p_j$ ($\f_j$). In this 
extension, several parameter regions previously discussed can be simultaneously realized 
with the neutrino portal interactions of different generations.

Now let us provide a natural realization of the UV model \Eq{nUV} for our relevant parameter ranges where $m_\f$ and $M$ are sufficiently small. A light scalar $\f$ suggests a naturalness problem.
One of the solutions to this problem\footnote{Alternatively, this may indicate that a SUSY extension of the SM 
has a SUSY breaking soft scale around $\MEV$s in the $Z_2$ odd sector, while that in the SM sector is above 
$\TEV$ to survive the experimental constraints. 
A candidate is a gauge mediation scenario \cite{Giudice:1998bp}, where sparticles charged under the SM 
gauge group gain weight via gauge interaction, while a singlet scalar acquires a highly suppressed mass 
either from higher order correction or the gravity effects.} is to identify $\f$ as a pseudo-Nambu-Goldstone 
boson.
Now consider the spontaneously breaking of an approximate $\SU(2)\times \U(1)$ global symmetry to $\U(1)$ 
by some non-perturbative effect in analogy with the chiral symmetry breaking in QCD. 
If all the explicit breaking terms of $\SU(2)\times \U(1)$ can be identified as spurions with even charges under the residual $\U(1)$, this 
residual symmetry contains an exact $Z_2$ symmetry.
The $\U(1)$ charged pion, say $\pi_{+}$, is $Z_2$ odd and contains $\f$ as $\pi_{+}={\f+i \tl{\f} \o \sqrt{2}}$.
This possibility not only explains the smallness of $m_\f$, but also allows a rather small decay constant, $f$, for 
the composite scalar $\f$, like the pion decay constant in QCD. 

To be concrete, let us consider the following non-linear realized Lagrangian for pions,
\beq
\laq{totUV}
{\cal L}_{UV}= {\cal L}_{sym}+{\cal L}_{exb}
\ebq
\laq{sym}
{\cal L}_{sym}= \ol{\vec{N}}  \ol{\s}^\m \partial_\m  \vec{N}-\vev{\vec{\F}}\cdot e^{i{\pi^a \s_a\o 2f}} \cdot \vec{N} S+h.c., 
\ebq
\laq{exb}
{\cal L}_{exb}=-{M_\p \o 2}\p \p
-{ m_\f^2 \o 2}\f^2-{ M_N \o 2} NN -{\tl{m}^2\o2} \tl{\f}^2 -{{m_0^2 \o 2}\pi_0^2}-y \f_H \cdot LN+h.c.
\eeq
Here, ${\cal L}_{sym}$ is $\SU(2) \times \U(1)$ symmetric Lagrangian, while terms, which explicitly break $\SU(2) \times \U(1)$, are collected in ${\cal L}_{exb}$; $ \vec{N}= \({N, \p}\)$ is a matter doublet with $\U(1)$ charge $-{1/2}$ and lepton number $-$1, while the 
fermion $S$ carries a lepton number 1; $\vev{\vec{\F}}=(\tl{M}, 0)$ is the VEV of an $\SU(2)$ doublet 
operator with $\U(1)$ charge $-{1/2}$, and the second term of \Eq{sym} turns out 
to be the second and third terms in \Eq{nUV}.

The mass parameters $M_\p, m_\f,
\tl{m},m_0$ are the explicit breaking terms of the $\SU(2)\times \U(1)$ symmetry, and can be smaller than $\tl{M}$ and $f$ naturally. 
In particular, the unbroken $\U(1)$ is explicitly broken down to exact $Z_2$ symmetry by $M_\p$ and $\tl{m}^2 - m_\f^2$. 
Since $\p$ can be also charged under lepton number instead of $\p$, the 1loop neutrino mass is suppressed by an additional factor of $\tl{m}^2-m_\f^2\o \tl{m}^2+m_\f^2$ and could reduce the tuning between $M_\p$ and $ m_\f$ to satisfy the neutrino mass constraint. 

In this model, with these additional light particles which are assumed to be lighter than the Higgs boson,
the testability in the future lepton colliders is even increased.
 Although the neutrino mass constraint is alleviated and $M_\p$ can deviate from $m_{\tl{\f}} \simeq m_{\f}$, 
for a given values of the mass and the cross section for dark matter-dark matter (neutrino-(anti)neutrino), the increase of $\max{( M_\p, m_{\tl{\f}}\simeq m_{\f})}$ leads to the increase of the neutrino portal coupling $1/M$. 
Thus, the Higgs invisible decay rate is even enhanced for the regions of thermal dark matter and affecting the propagation of the cosmic-ray neutrino. \\

\section*{Acknowledgement}
I would like to thank Adam Falkowski, Fapeng Huang, and Hao Zhang for collaboration at an early stage of this work. 
I also thank Hiroyuki Ishida and Yingnan Mao for useful discussions and thank Hiromasa Takaura for carefully reading the manuscript. 
Moreover, I thank the referee for carefully checking the calculations as well as pointing out typos and confusing statements.

\end{document}